\documentclass{PoS}

\usepackage{graphicx}
\usepackage{subfig}

\title{Thermal effective potential for the Polyakov loop to higher loop order}

\ShortTitle{Thermal effective potential for the Polyakov}

        \author{\speaker{Hiromichi Nishimura}\thanks{HN is supported by the RIKEN Special Postdoctoral Researchers Program.}\\
   RIKEN-BNL Research Center, Brookhaven National Laboratory, Upton, New York 11973, USA\\
 E-mail: \email{hnishimura@bnl.gov}}

\author{Chris Korthals-Altes\\
 NIKHEF Theory Group
Science Park 105
1098XG Amsterdam
The Netherlands \\
Centre Physique Theorique
Campus de Luminy, Case 907
163 Avenue de Luminy
13288 Marseille Cedex 9, France\\
        E-mail: \email{chrisaltes@gmail.com}}

          \author{Robert D. Pisarski\\
        Department of Physics, Brookhaven National Laboratory, Upton, NY 11973\\
        E-mail: \email{rob.pisarski@gmail.com}}

       \author{Vladimir Skokov\\
       Department of Physics, North Carolina State University, Raleigh, North Carolina 27695, USA\\
   RIKEN-BNL Research Center, Brookhaven National Laboratory, Upton, New York 11973, USA\\
        E-mail: \email{vskokov@ncsu.edu}}

\abstract{
  This is a progress report on the calculation of the effective potential for 
  the Polyakov loop in $SU(N)$ pure gauge theory beyond two-loop order. 
  We introduce a new approach using the Poisson resummation formula to compute sum-integrals with the holonomy. 
  We show partial results for the free energy at order $g^3$ and $g^4$.
}

\FullConference{XIII Quark Confinement and the Hadron Spectrum - Confinement2018\\
		31 July - 6 August 2018\\
		Maynooth University, Ireland}

\begin{document}

\section{Introduction}
Our goal is to compute the effective potential for the Polyakov loop in pure $SU(N)$ Yang-Mills theory beyond two-loop order.  The Polyakov loop is the order parameter for the deconfinement phase transition, for a review see~\cite{Fukushima:2017csk}. At two-loop order, the minimum of the potential corresponds to completely broken $Z(N)$ center symmetry.  The pressure up to two loops is thus given by perturbative Yang-Mills theory without the holonomy (the Wilson line).  At three loop order and beyond, the effective potential is not known. 

The Polyakov loop is defined as a Wilson loop that winds in the temporal direction:
\begin{equation}
\mathrm{tr}\,  P(\mathbf{x})  = \mathrm{tr}\,  \mathcal{P} \exp \left[ i g \int^{\beta}_{0} d \tau A_4 ( \mathbf{x}, \tau) \right],
\end{equation}
where $\beta = 1/T$ is the inverse temperature.
One way to construct the effective potential $V$ for the Polyakov loop is to compute the path integral with the constraints \cite{KorthalsAltes:1993ca,Dumitru:2013xna}:
\begin{equation}
e^{- \beta \mathcal{V} V(L_n)} 
= 
\int DA \left[ \prod^{N-1}_{m=1} \delta \left(L_m - \frac{1}{\mathcal{V}}\int d^3 x \frac{1}{N}  \mathrm{tr}\,  P^m(\mathbf{x}) \right) \right]e^{-S_{\rm YM}(A)},
\end{equation}
where $\mathcal{V}$ is the three volume. The delta functions set  
 the spatial average of the $m$-th winding of the Polyakov loop to be some value $L_m$. For $SU(N)$, $N-1$ independent constraints are required. 
The effective potential then consists of two parts:
\begin{equation}
V = V_{\rm free} + V_{\rm insert}. 
\label{V_total}
\end{equation}
$V_{\rm free}$ is the usual free energy in the presence of the background field $\bar{A}_{\mu} = \delta _{\mu 4} \bar{A}_{4}$.  We choose the background field to be constant and diagonal:
\begin{equation}
g \beta \bar{A}^a_4 = \theta_a = \theta_{ij} = \theta_i - \theta_j , 
\end{equation}
where $a,b= 1 \dots N^2-1$ and $i,j = 1 \dots N$ are the indices for the adjoint and fundamental representations in $SU(N)$, respectively.
The other term in the effective potential, $V_{\rm insert}$,  comes from the delta function constraints.  
It is known that $V_{\rm free}$ depends on the gauge fixing parameter $\xi$ at two-loop order \cite{Enqvist:1990ae}.  The full combined effective potential, $V_{\rm free}+V_{\rm insert}$, is gauge invariant by construction. 
The gauge invariance at order $g^2$ was explicitly shown in \cite{KorthalsAltes:1993ca,Dumitru:2013xna}\footnote{There is an alternative approach to obtain the gauge invariant result at order $g^2$ as done in \cite{Belyaev:1991gh}.}. 

The leading order $V$ is the Gross-Pisarski-Yaffe-Weiss potential \cite{Gross:1980br,Weiss:1980rj}. The next-to-leading order turns out to be proportional to the GPY-Weiss potential \cite{KorthalsAltes:1993ca,Dumitru:2013xna,Belyaev:1991gh}. 
The 
 effective potential up to order $g^2$ modulo loop-independent  terms  can be written as
\begin{equation}
V = - \frac{N^2  T^4}{\pi^2} \left(1 - \frac{5 g^2 N}{16 \pi^2} \right)  \sum_{n \neq 0} \frac{L_n L_{-n} }{n^4}, 
\end{equation}
where 
\begin{equation}
L_n = \frac{1}{N} \sum^{N}_{i =1} e^{ i n \theta_i}
\end{equation}
with $\sum^N_i \theta_i =0$. 
It is remarkable that there are only two traces up to order $g^2$. A naive power counting predicts three-trace terms  at $g^2$ order.  
Because the order of phase transition at large $N$ depends on the structure of the potential in terms of the Polyakov loops \cite{Nishimura:2017crr}, computing at the three-loop order has some theoretical interests as well.  



\section{Evaluation of sum-integrals with the holonomy}
\label{sec:Poisson}
In this section, we establish a new technique to compute sum-integrals with the holonomy.  
In the calculation of the Polyakov loop effective potential at higher loop order,
there are many sum-integrals which we find difficult to compute using the standard methods, such as performing the Matsubara sum directly or an analytic continuation of the momentum in the temporal direction $p_4$ \cite{Kapusta:1979fh}.  We show that the new method is more suited for our purpose using some examples, which are relevant for the three-loop calculations.  

\subsection{Poisson resummation}
Consider a periodic function, $f(\theta + 2 \pi) = f(\theta)$, of the form
\begin{equation}
f(\theta) = \sum_{n \in \mathbb{Z}} F(2 \pi n + \theta) , 
\end{equation}
where $\theta$ is real. 
The Fourier series of $f(\theta)$ is
\begin{equation}
f(\theta) = \sum_{m \in \mathbb{Z}} \tilde{f}_m e^{i m \theta} .
\end{equation}
The Fourier coefficients are given by
\begin{eqnarray}
\tilde{f}_m 
&=&
\int^{2 \pi}_0 \frac{d \theta}{2 \pi} e^{-im \theta} f(\theta)
\\
&=&
\int^{\infty}_{-\infty} \frac{d \theta}{2 \pi} e^{-im \theta} F(\theta) .
\label{Poisson_FourierCoefficient}
\end{eqnarray}
This is a version of the Poisson resummation, which was used e.g. in \cite{Poppitz:2011wy,Nishimura:2011md}.

In perturbative finite-temperature field theory with the nontrivial holonomy, $f(\theta)$ arises naturally from Feynman diagrams, where $F (p_4 /T)$ is obtained after the spatial momentum integral. The Poisson duality in this context is the duality between the Matsubara mode $n$ in $p_4$ and the $A_4$ charge $m$ in the $m$-th winding of the Polyakov loop. This duality provides a mathematical connection between the effective potential in terms of the Polyakov loops $L_m$ and the one in terms of  the eigenvalues $\theta$.  The other advantage of this method is that the sum is  replaced by the integral as shown in Eq.~(\ref{Poisson_FourierCoefficient}), and this makes it more analytically tractable as we show below. 

The zero winding term $m=0$  is nothing but the zero-temperature contribution.  As in the method of analytic continuation of $p_4$, the Poisson resummation can naturally separate the zero-temperature and finite-temperature contributions.  Another advantage of the Poisson resummation is that the order of the Matsubara sum and the spatial momentum integral can be interchanged, while the method of analytic continuation requires to perform the Matsubara sum first.   At higher loop order, doing the spatial momentum before the Matsubara sum is essential in order to factorize two momentum as argued in \cite{Arnold:1994ps}. 

\subsection{Examples}
We compute a sum-integral of the form
\begin{equation}
f^{\left( u,v \right)} (\theta^a) = \int_{p} \frac{1}{ \left( p^a \right)^{2u} \left(  p^a_4 \right)^{v}} , 
\end{equation}
where $u, v$ are real numbers and 
\begin{equation}
\int_p \equiv \sum_{n_p \in \mathbb{Z}} \int \frac{d^{d} p}{\left( 2 \pi \right)^{d}}. 
\end{equation}
The four-momentum is defined as
\begin{equation}
p^a_4= \left( 2 \pi n_p  + \theta^a \right)T 
\;\;\;\;\;
\mbox{and}
\;\;\;\;\;
\left( p^a \right)^2 = \left( p^a_4 \right)^2 + p^2 .
\end{equation}
We use the Poisson resummation formula to get
\begin{equation}
f^{\left( u,v \right)} (\theta^a)  =  T \sum_{n_p \in \mathbb{Z}} F(2 \pi n_p + \theta^a) = T \sum_{m_p \in \mathbb{Z}} \tilde{f}^{\left( u,v \right)}_{m_p} e^{i m_p \theta^a}. 
\end{equation}
We evaluate $\tilde{f}^{\left( u,v \right)}_{m_p}$ to compute the sum-integral.

First we perform the momentum integral using the dimensional regularization, 
\begin{eqnarray}
\mu^{2 \epsilon} \int \frac{d^{d} p}{ \left(2 \pi \right)^d} \frac{1}{ \left( p^a \right)^{2u} }
=
 \mu^{2 \epsilon} \frac{ \Gamma(u-\frac{d}{2})}{ 2^d \pi^{\frac{d}{2}} \Gamma(u)} 
\left| p^a_4 \right|^{d -2 u }
\label{spatial_momentum_integral_DR}
\end{eqnarray}
where $d= 3- 2 \epsilon$.  $\mu$ is the scale in the dimensional regularization.  We thus have
\begin{eqnarray}
F(\theta)
=
\frac{c_{d,u}}{T^v} \frac{\left|  \theta \right|^{d-2u}}{ \theta^v}
\label{F_theta_1}
\end{eqnarray}
where
\begin{equation}
c_{d,u} = \frac{ \Gamma(u-\frac{d}{2})}{ 2^d \pi^{\frac{d}{2}} \Gamma(u)} T^{3-2u} \left( \frac{ \mu}{T} \right)^{2 \epsilon} . 
\end{equation}

Second we compute $\tilde{f}^{\left( u,v \right)}_{m_p}$ using Eq.~(\ref{Poisson_FourierCoefficient}):
\begin{eqnarray}
\tilde{f}^{\left( u,v \right)}_{m_p} 
&=&   
\frac{c_{d,u}}{T^v}  \int^{\infty}_{-\infty} \frac{d\theta}{2 \pi}\frac{\left|  \theta \right|^{d-2u}}{ \theta^v} e^{-i m_p \theta}
\\
&=&
\frac{c_{d,u}}{T^v} \int^{\infty}_{0} \frac{d \theta}{2\pi} \theta^{d-2u-v} \left( e^{-i m_p \theta} + e^{i \pi v} e^{i m_p \theta}\right) .
\end{eqnarray}
For $m_p \neq 0$ we use the identity
\begin{equation}
\int^{\infty}_{0} d  \theta \, \theta^{-1+ \alpha} 
e^{ i z \theta} =
\frac{e^{i \pi \alpha}\Gamma(\alpha)}{ z^{\alpha}} 
\label{Fourier_identity1}
\end{equation}
where $0<\mathrm{Re} (\alpha) < 1$ and $z$ is real and nonzero.  This gives
\begin{equation}
\tilde{f}^{\left( u,v \right)}_{m_p}
=
\frac{c_{d,u} \, \Gamma(d-2u-v+1)}{2 \pi \, T^v \, m^{d-2u-v+1}_p} 
\left( e^{ - i \frac{\pi}{2} \left( d-2u-v+1 \right) } + e^{ i \frac{\pi}{2} \left( d-2u+v+1 \right)} \right).
\end{equation}

Performing small-$\epsilon$ expansion for $u=1$ and $v=0,1,2$, we get
\begin{eqnarray}
\left\{ 
\tilde{f}^{\left(1,0 \right)}_{m_p}, 
\tilde{f}^{\left(1,1 \right)}_{m_p},
\tilde{f}^{\left(1,2 \right)}_{m_p}
\right\}
&=& 
\left\{
\frac{T}{4 m^2_p \pi^2}, 
\frac{i}{4 m_p \pi^2},
\frac{1}{8 \pi^2 T \epsilon} + \frac{\tilde{c}  +  \log(m^2_p) }{8 \pi^2 T}
\right\} .
\label{ftilde_20}
\end{eqnarray}
In this equation, $m_p\neq 0$ and $\tilde{c} =2 + \gamma + \log( \pi \mu^2 /T^2)$.  For the sake of brevity we will not discuss the zero-temperature contributions $m_p=0$ in this proceedings. 

\section{One-loop self-energy with the holonomy}
\label{sec:SelfEnergy}
The one-loop self-energy with the incoming momentum $p_a$ in the Feynman gauge ($\xi=1$) can be written as
\begin{equation}
\Pi^{\mu \nu}_{ad} (p_a)
=
g^2 f_{abc} f_{\bar{b}\bar{c} \bar{d}}
\Pi^{\mu \nu} (p_a, \theta_b, \theta_c)
\label{SelfEnergy}
\end{equation}
where
\begin{equation}
\Pi^{\mu \nu} (p_a, \theta_b, \theta_c)
\equiv
 \int_q
\frac{1}{q^2_b r^2_c} 
\left\{ 
\left[\left(D-2 \right) q^2_b -2 p^2_a   \right] g^{\mu\nu} 
-2 \left(D-2\right) q^{\mu}_b q^{\nu}_b + \left( 1 + \frac{D}{2} \right) p^{\mu}_a p^{\nu}_a
\right\}
\end{equation}
with $r_c = -p_a -q_b$.  Here $D=d+1$ is the space-time dimension.  In the case of the trivial holonomy, this reduces to the result of \cite{Elze:1987rh}. 

We note that the structure constant that appears in the self-energy $f_{abc} f_{\bar{b}\bar{c} \bar{d}}$ is invariant under the exchange of $b \leftrightarrow c$.  We can therefore use the following manipulation:
\begin{equation}
f_{abc} f_{\bar{b} \bar{c} \bar{d}} \int_q h(q_b,r_c) 
= 
f_{abc} f_{\bar{b} \bar{c} \bar{d}} \int_{qr} \delta(p+q+r) h(q_b,r_c) 
=
f_{abc} f_{\bar{b} \bar{c} \bar{d}} \int_{q} h(r_c,q_b) , 
\label{trick_1}
\end{equation}
where we have used the change of variables $q_b \leftrightarrow r_c$.
Using Eq.~(\ref{trick_1}), the following identity can be proven
\begin{eqnarray}
f_{abc} f_{\bar{b} \bar{c} \bar{d}} \int_q \frac{q^{\mu}_b}{q^2_b r^2_c}
=
-\frac{1}{2}
f_{abc} f_{\bar{b} \bar{c} \bar{d}}
\int_{q} \frac{p^{\mu}_a}{q^2_b r^2_c} .
\label{trick_2}
\end{eqnarray}
This identity was used to obtain the expression given in Eq.~(\ref{SelfEnergy}).

Eq.~(\ref{trick_2}) can be also used  to express the self-energy in another form:
\begin{equation}
\Pi^{\mu \nu} (p_a, \theta_b, \theta_c)
=
\frac{D-2}{2} \bar{\Pi}^{\mu \nu} (p_a, \theta_b,  \theta_c)
- 2 
 \left(p^2_a g^{\mu \nu} - p^{\mu}_{a} p^{\nu}_{a} \right)
\int_q \frac{1}{q^2_b r^2_c}
\label{SelfEnergy_ver2}
\end{equation}
where
\begin{equation}
\bar{\Pi}^{\mu \nu} (p_a, \theta_b, \theta_c)
\equiv
2 g^{\mu \nu} \int_q \frac{1 }{r^2_c}
-
\int_q \frac{\left( 2 q_b + p_a\right)^{\mu} \left( 2 q_b + p_a\right)^{\nu}}{q^2_b r^2_c} 
.
\end{equation}
This particular decomposition was used by Arnold and Zhai to compute the $g^4$ contribution to the pressure in the case of trivial holonomy. Without the holonomy both $\Pi^{\mu \nu}$ and $\bar{\Pi}^{\mu \nu}$ are transverse.

In the presence of a nontrivial holonomy, $\bar{\Pi}^{\mu \nu}$ is no longer transverse. Using Eq.~(\ref{SelfEnergy_ver2}), we can compute the non-transversity of the self-energy explicitly:
\begin{eqnarray}
p^{\mu}_a \Pi^{\mu \nu}_{ad} (p_a)
&=&
-g^2 2 \left(D-2\right)
 f_{abc} f_{\bar{b}\bar{c} \bar{d}}
\int_q \frac{q^{\nu}_b}{q^2_b}
\\
&=&
-g^2  \frac{4\pi}{3} T^3 \left(D-2\right)
 f_{abc} f_{\bar{b}\bar{c} \bar{d}}
 B_3 \left( \frac{\theta_b}{2 \pi} \right)
 \delta^{\nu 4}
 \\
 &=&
\frac{4 \pi}{3} g^2 T^3 \sum_{k} \left[B_3 \left(\frac{\theta_{ik}}{2\pi} \right) - B_3 \left(\frac{\theta_{jk}}{2\pi} \right) \right]   \delta^{\nu 4} 
\label{KA_Identity}
\end{eqnarray}
where we completed the square and used Eqs.~(\ref{trick_1}) and (\ref{trick_2}) to get the first line. $B_3(x)$ is the third Bernoulli polynomial. In the last line, we used $a=ij$ and $D=4$.   This identity was first derived in \cite{KorthalsAltes:1993ca} using the BRS symmetry; see also \cite{Hidaka:2009hs} where this identity was discussed in a different context.
 The non-transversity of the self-energy  makes higher-loop calculations more complicated but plays  an important role for the gauge invariance of the effective potential.

\subsection{Static limit}
The self-energy in the static limit ($p_4 =0$ and $\mathbf{p} \rightarrow 0$) is required for the $g^3$ computation in the next section. From Eq.~(\ref{SelfEnergy_ver2}), we obtain  
\begin{eqnarray}
\Pi^{44}(\theta_a, \theta_b, \theta_c)
&=&
2  \int_q \frac{1}{q^2_c} - \int_q \frac{\left(2 q^b_4 + \theta^a T \right)^2}{q^2_b q^2_{\bar{c}}} 
\label{SelfEnergy_44_SL}
\\
\Pi^{xx} (\theta_a, \theta_b, \theta_c)
&=&
2 \int_q \frac{1}{q^2_c} - 4 \int_q \frac{q^2_x}{q^2_b q^2_{\bar{c}}} - 2 \left( \theta^a T\right)^2 \int_q \frac{1}{q^2_b q^2_{\bar{c}}}
\label{SelfEnergy_xx_SL}
\end{eqnarray}
and $\Pi^{\mu \nu}_{ad}(\theta^a)=0$ when $\mu \neq \nu$.
For the trivial holonomy $\theta = 0$, we obtain the usual expression:
\begin{equation}
\Pi^{44}_{ad} = \frac{1}{3} N g^2 T^2   \delta_{ad} = m^2_D \delta_{ad}
\label{SelfEnergy_Debye}
\end{equation}
where $m_D$ is the Debye mass.
In general we can evaluate the eigenvalues of the self-energy with the nontrivial holonomy in the static limit by computing Eqs.~(\ref{SelfEnergy_44_SL}) and (\ref{SelfEnergy_xx_SL}) and then diagonalizing the matrix in Eq.~(\ref{SelfEnergy}). 

The self-energies for the charged (off-diagonal) gluons with $\theta_a = \theta_{ij}$, $i \neq j$,  do not mix with other states.
Therefore the diagonalization is not required for the charged gluons:
\begin{equation}
 \Pi^{\mu \nu}_{ij, ji} = \delta^{\mu \nu}   \Lambda^{ij}_{\mu}.
 \label{MassEV_Off}
\end{equation}
Using the identity (\ref{KA_Identity}), we obtain
\begin{eqnarray}
\Lambda^{ij}_{\mu} 
&=&  
\delta_{\mu4}
\frac{4 \pi}{3} g^2 T^2 \sum^N_{k=1} \frac{B_3\left(\frac{\theta_{ik}}{2\pi}\right) - B_3\left(\frac{\theta_{jk}}{2\pi}\right)}{\theta_{ij}} . 
\label{Lambda4_offdiagonal}
\end{eqnarray}

The self-energies for the neutral (diagonal) gluons with $\theta_{a}=0$, on the other hand, can mix with other diagonal gluons.  
In this case we have $q_{\bar{c}} =q_b$. The self-energies for the neutral gluons are
\begin{eqnarray}
\Pi^{44}(0,\theta_b,\theta_c)
&=&
4   \int_q  \frac{1}{q^2_b}
=
2  T^2  B_2\left(\frac{\theta_b}{2 \pi} \right)
\\
 \Pi^{xx}(0,\theta_b,\theta_c)
&=&
0
\end{eqnarray}
where we have used the following identity:
\begin{equation}
\int d^d p\frac{p^2}{\left( p^2 + m^2 \right)^k} 
=
\frac{d \,  \Gamma(k-1)}{2 \, \Gamma(k)} \int d^d p \frac{1}{\left( p^2 + m^2 \right)^{k-1}} .
\end{equation}
The mass matrix for the neutral gluons then becomes
\begin{equation}
\Pi^{44}_{ad} = 2 g^2 T^2 f_{abc} f_{\bar{b} \bar{c} \bar{d}} B_2 \left(\frac{\theta_b}{2\pi} \right) .
\end{equation}
Because of the off-diagonal elements, it is usually not straightforward to diagonalize the mass matrix analytically except for $N=2$ or $N=3$. Nevertheless as any symmetric matrix, it can be diagonalized to yield $N-1$ mass eigenvalues
\begin{equation}
 \Pi^{44}_{ab} (0) \rightarrow 
   \Pi^{44}_{ab} (0)= \mbox{diag} 
 \left(\Lambda^{1}_4, \Lambda^{2}_4, \dots, \Lambda^{N-1}_4 \right)
 \label{m_neutral}
 \end{equation}
for the $N-1$ neutral gluons in this new basis.  We use the same indices  $a,b$ for the new basis below.  

\section{Free energy at order $g^3$}
\label{sec:FreeEnergy_g3}
As well known in perturbative Yang-Mills theory without the holonomy, due to resummation of the infrared divergencies  present beyond the two-loop order $g^2$ in ring diagrams, the next order is not $g^4$ but $g^3$ \cite{Kapusta:1979fh}.  
Similar divergence can be expected with the holonomy. Consider 
\begin{equation}
V^{(\rm ring)}
=
-
\frac{1}{2}\mathrm{tr} \int_p \sum^{\infty}_{k=2}\frac{1}{k} \left[ -\Pi^{\alpha \gamma}_{ac} (p)D^{\gamma \beta}_{c b} (p) \right]^k  .
\end{equation}
The potential infrared divergence comes from the zero mode. In order to perform the trace operation, we need to diagonalize the $N^2 -1 $ matrices $\Pi_{ab}$ in the static limit, which are $4 \times 4$ matrices.  
The neutral gluon propagator mixes with other neutral gluon propagator, while the charged gluons do not mix as argued in the previous section.

We first consider the neutral gluons.  In this case, $\theta_a$ in the propagator is zero, so it gives a usual resummation of ring diagrams. The only modification is that we have to use the new basis where the self-energies are diagonal. We therefore have
\begin{eqnarray}
V^{(\rm ring)}_{\rm n.g.}
&\sim&
\frac{1}{2} T \sum^{N-1}_{d=1} \int \frac{d^3p}{ \left(2 \pi \right)^3} \left[ \ln \left(1+ \frac{\Lambda^d_4}{ p^2}  \right) - \frac{\Lambda^d_4 }{ p^2} \right]
\end{eqnarray}
where $\Lambda^d_4$ is the eigenvalue of the self-energy for the neutral gluons $\Pi^{44}_{ab}$ (\ref{m_neutral}). 
Performing the momentum integral, we obtain the $g^3$ contribution from the neutral gluons 
\begin{equation}
V^{\left( 3 \right)}_{\rm n.g.} 
=
-\frac{T^4}{12\pi} \sum^{N-1}_{d=1} \left( \frac{\Lambda^d_4}{T^2} \right)^{3/2} .
\label{Gamma3_ng}
\end{equation}
A more explicit form of $V^{\left( 3 \right)}_{\rm n.g.}$ for $SU(3)$ is given in \cite{Enqvist:1990ae,Giovannangeli:2002uv}. 

For the charged gluons, the self-energies are already diagonal.  Taking only the zero mode, 
\begin{eqnarray}
V^{(\rm ring)}_{\rm c.g.}
&\sim&
\frac{1}{2} T \sum_{\alpha, i \neq j} \int \frac{d^3p}{ \left(2 \pi \right)^3} \left[ \ln \left(1+ \frac{\Lambda^{ij}_\alpha}{\theta^2_{ij} T^2+ p^2}  \right) - \frac{\Lambda^{ij}_\alpha }{ \theta^2_{ij} T^2 + p^2} \right],
\end{eqnarray}
and performing the momentum integral, we get
\begin{eqnarray}
 V^{\left(3 \right)}_{\rm c.g.}
&=&
-T \sum_{\alpha,i \neq j} \left[
\frac{1}{12 \pi} \left\{ \left(\theta^2_{ij} T^2 + \Lambda^{ij}_{\alpha}  \right)^{3/2} - \left| \theta_{ij} T \right|^3 \right\}
- \frac{1}{8 \pi}   \Lambda^{ij}_{\alpha}  \left| \theta_{ij} T \right| \right].
\label{Ring_cg}
\end{eqnarray}

The total $g^3$ contribution to the free energy is  the sum of the two terms 
\begin{equation}
V^{\left(3 \right)}_{\rm free} = V^{\left(3 \right)}_{\rm n.g.} + V^{\left(3 \right)}_{\rm c.g.} . 
\end{equation}
If $\theta=0$,  $\Lambda^a_4= m^2_D$, and we have $V^{\left( 3\right)}_{\rm free} = -\left( N^2-1 \right)T m^3_D/ (12 \pi)$, which is the known perturbative result  without the holonomy.
All $N^2-1$ gluons contribute to the $g^3$ order. 
On the other hand, when $\theta \sim \mathcal{O}(1)$, only the first term $V^{\left(3 \right)}_{\rm n.g.}$ accounting for
 $N-1$ neutral gluons  contributes, as the remaining $N(N-1)$  gluons do not display an IR divergence owing to an effective mass proportional to $T \theta$.  In  \cite{Enqvist:1990ae} it was noticed that the eigenvalues of the self-energy become negative for sufficiently large $\theta$ near the confined phase.  The window of this instability becomes narrower as $N$ becomes larger \cite{Giovannangeli:2002uv}. 

\section{Free energy at order $g^4$}
\label{sec:FreeEnergy_g4}

\begin{figure}[t]\centering
\begin{minipage}{.13\textwidth}
  \subfloat[$I^{\left(3 a \right)}$]
  {
  \includegraphics[width=\textwidth]{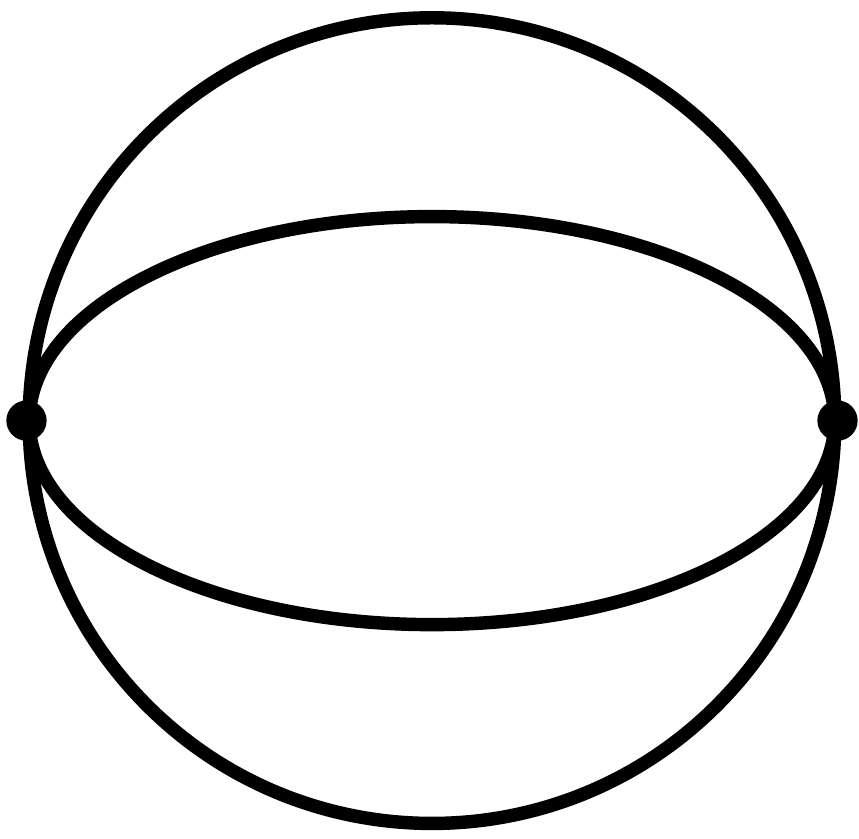}
  \label{I_3a}
  }
\end{minipage}
\begin{minipage}{.13\textwidth}
  \subfloat[$I^{\left(3 b \right)}$]
  {
  \includegraphics[width=\textwidth]{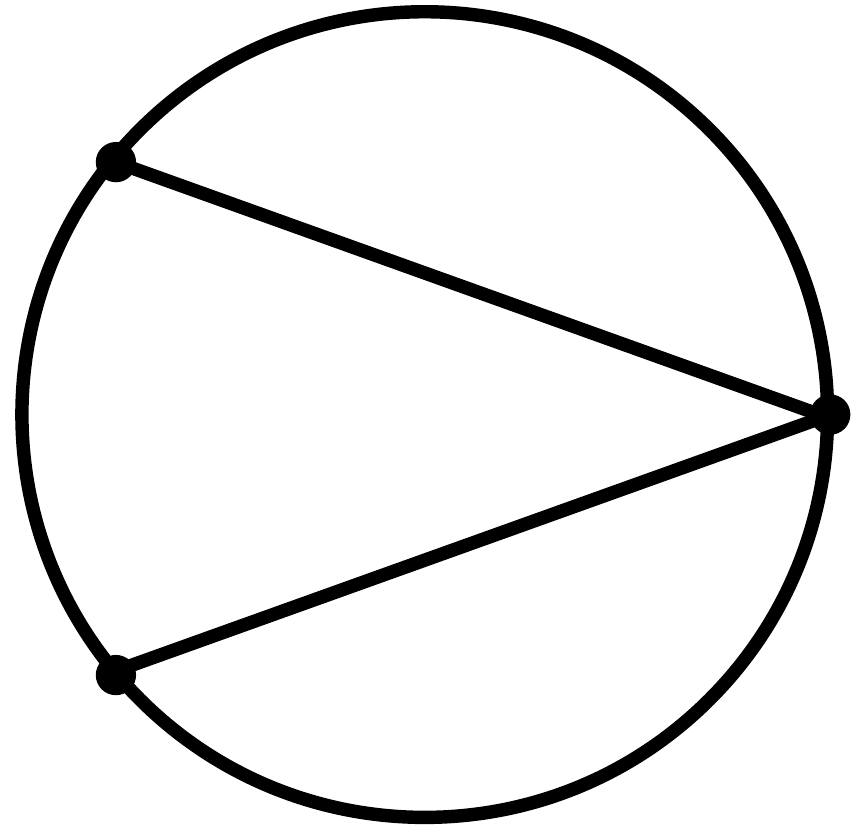}
  \label{I_3b}
  }
\end{minipage}
\begin{minipage}{.13\textwidth}
  \subfloat[$I^{\left(3 c \right)}$]
  {
  \includegraphics[width=\textwidth]{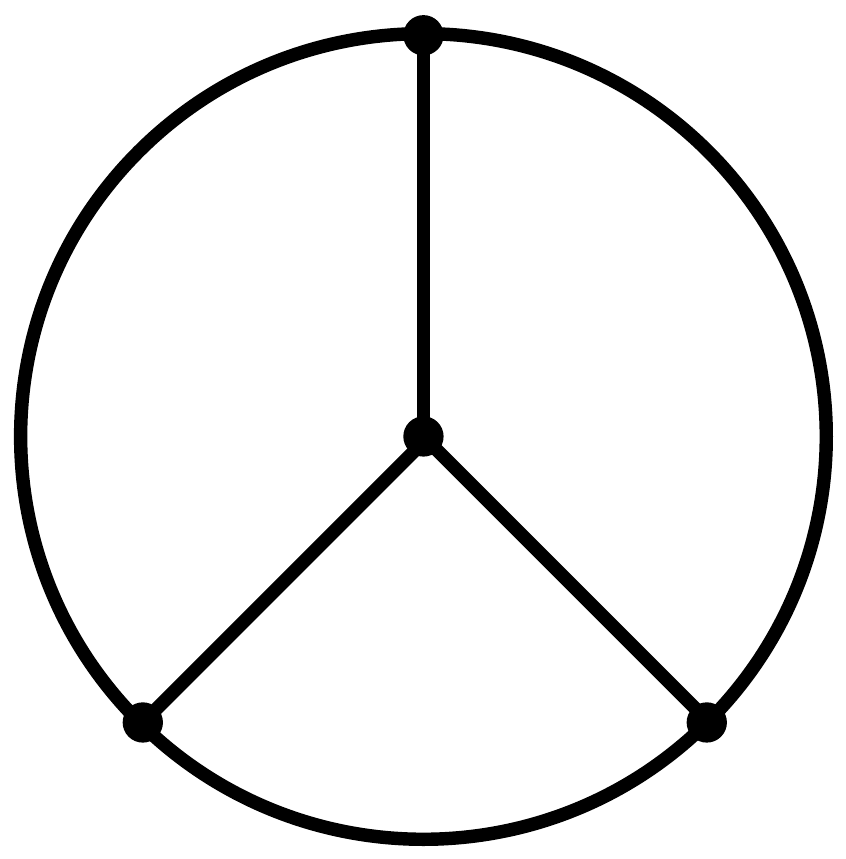}
  \label{I_3c}
  }
\end{minipage}
\begin{minipage}{.13\textwidth}
  \subfloat[$I^{\left(3 d \right)}$]
  {
  \includegraphics[width=\textwidth]{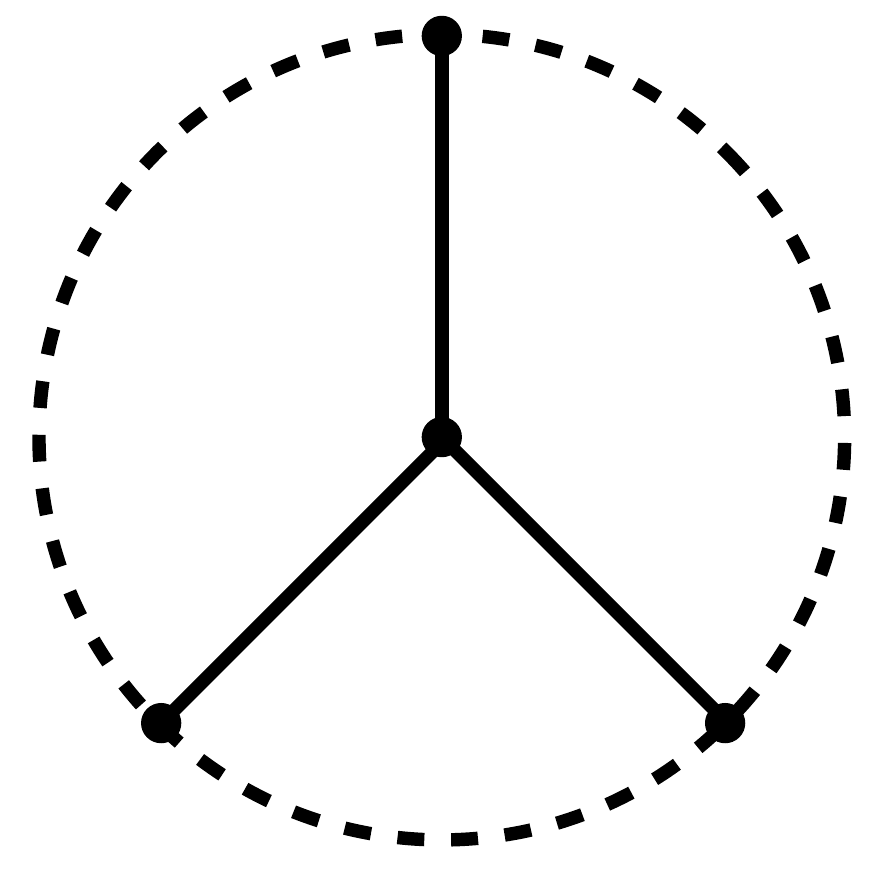}
  \label{I_3d}
  }
\end{minipage}
\begin{minipage}{.13\textwidth}
  \subfloat[$I^{\left(3 e \right)}$]
  {
  \includegraphics[width=\textwidth]{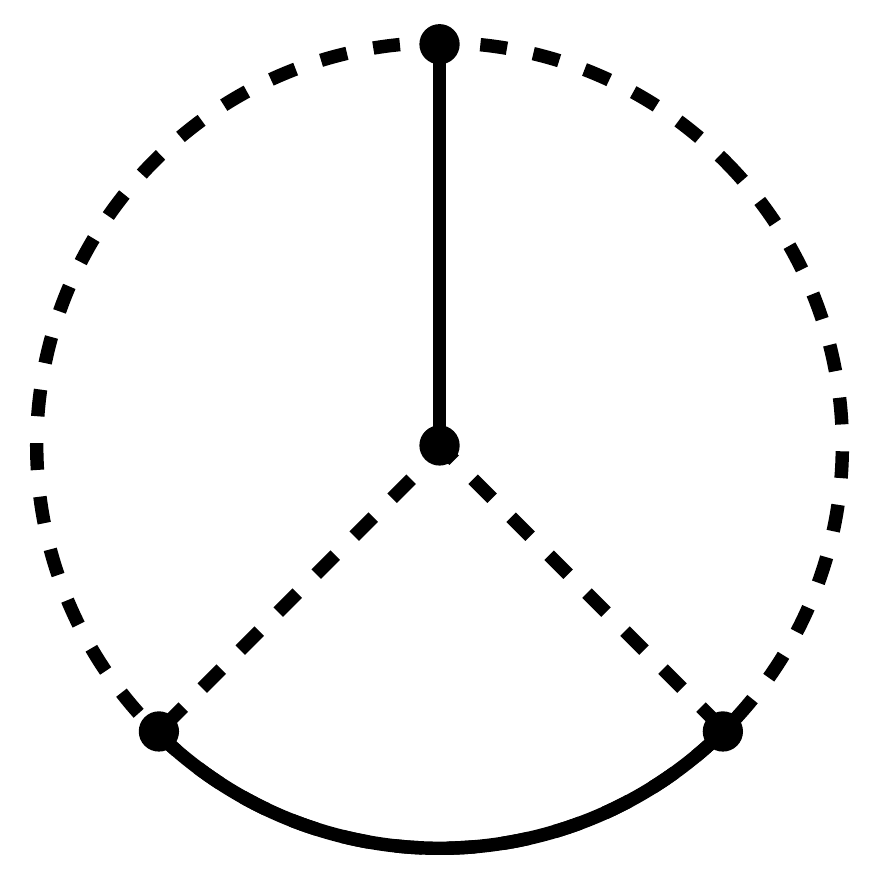}
  \label{I_3e}
  }
\end{minipage}
\begin{minipage}{.13\textwidth}
  \subfloat[$I^{\left(3 f \right)}$]
  {
  \includegraphics[width=\textwidth]{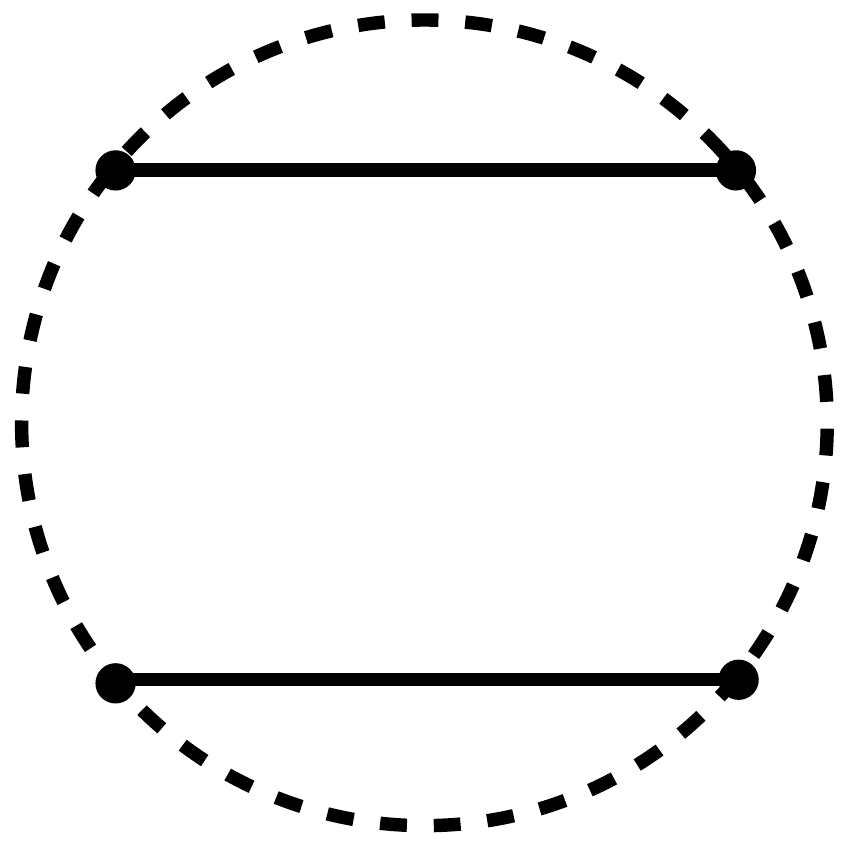}
  \label{I_3f}
  }
\end{minipage}
\begin{minipage}{.1075\textwidth}
  \subfloat[$I^{\left(3 g \right)}$]
  {
  \includegraphics[width=\textwidth]{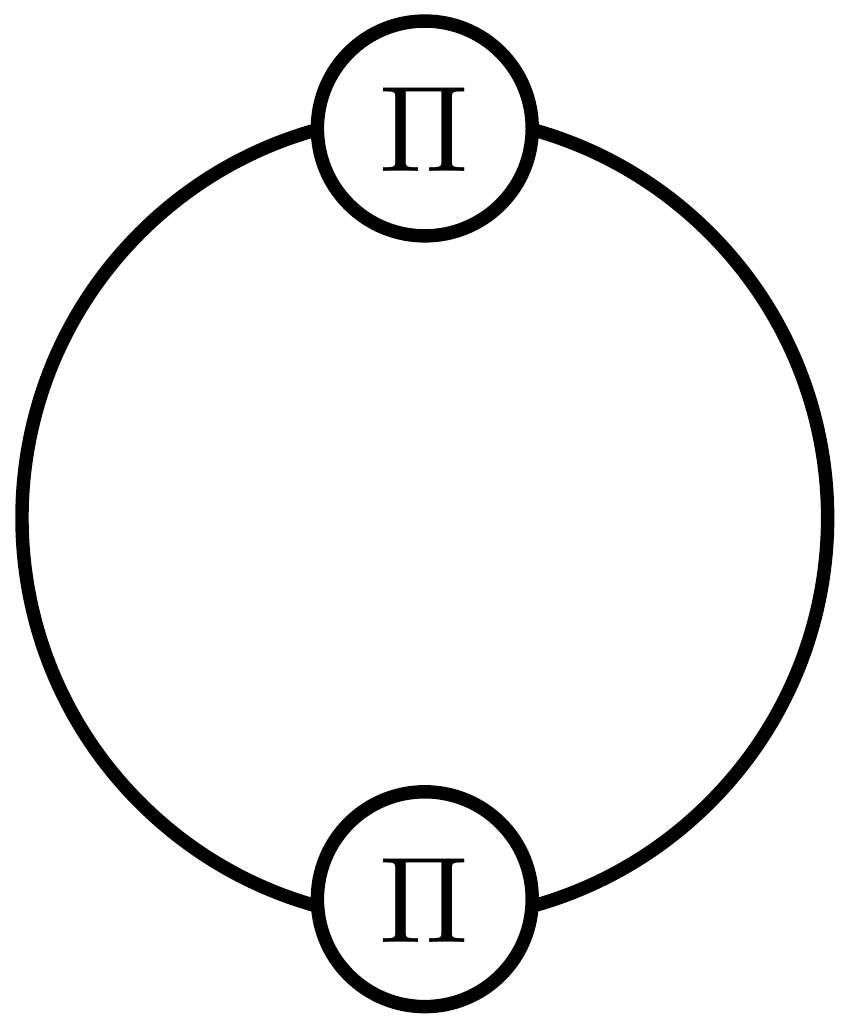}
  \label{I_3g}
  }
\end{minipage}
\caption{
  Three-loop free energy diagrams. The solid and dashed lines are the gluon and ghost propagators, respectively.  $\Pi$ is the one-loop self-energy. 
}
\label{3LoopDiagrams}
\end{figure}

The three-loop free energy is given by the sum of all the Feynman diagrams shown in Fig.~\ref{3LoopDiagrams}. Including the symmetry factor and the negative sign due to the ghost loop, we have
\begin{equation}
- V^{\left( 4 \right)}_{\rm free}
= 
\frac{1}{48}I^{\left( 3a \right)} +\frac{1}{8}I^{\left( 3b \right)} + \frac{1}{24}I^{\left( 3c \right)} 
- \frac{1}{3}I^{\left( 3d \right)} - \frac{1}{4}I^{\left( 3e \right)} -\frac{1}{2}I^{\left( 3f \right)}+ \frac{1}{4}I^{\left( 3g \right)} .
\end{equation}
We now take the large-$N$ limit so that we can reduce the sum-integrals as follows.  Using the momentum and color conservation, each Feynman diagram can be written as the sum-integral in terms of three momenta, $p_a$ $q_b$ and $r_c$. We use the double line notation and define $a= ij, \; b=kl, \; c=mn $.
Using 
\begin{equation}
f_{ij,kl,mn} = \frac{i}{\sqrt{2}} \left( \delta_{il} \delta_{kn} \delta_{mj} - \delta_{in} \delta_{kj} \delta_{ml} \right)
\end{equation}
we can sum over the other color indices.  We then  complete the square, $p_a \cdot q_b 
= 
\frac{1}{2} p^2_a+ \frac{1}{2} q^2_b - \frac{1}{2} \left( p_a -q_b \right)^2$,
and take into account  the symmetry of the diagrams to simplify the numerators. 

It turns out that we can reduce the Feynman diagrams (3a)-(3f) into a linear combination of two master integrals.  
We summarize the results of the reduction for  the diagrams $I^{\left(3 a\right)} - I^{\left(3 f\right)}$:
\begin{eqnarray}
\left\{ I^{\left( 3a \right)}, I^{\left( 3b \right)}, I^{\left( 3c \right)}, I^{\left( 3d \right)}, I^{\left( 3e \right)}, I^{\left( 3f \right)} \right\}
&=&
\left\{
9 D \left(D-1 \right) I_2,
\left(D-1 \right) \left(D-16 \right) I_2 - 2 \left(D-1 \right)  \left( 2 D -5 \right) I_3,
\right.
\nonumber
\\
&&
\left.
\frac{27}{2} \left( D - 1 \right) I_2 +3 \left( 2 D -5 \right) I_3,
\frac{3}{4} I_3,
\frac{1}{4} I_{2} - \frac{1}{2} I_{3},
\frac{1}{4} I_2
\right\}
\end{eqnarray}
where the master integrals, $I_2$ and $I_3$, are given by
\begin{eqnarray}
I_2
&=& 
\sum_{ijkl} \int_{pqr} 
\frac{1}{\left( p_{ij} -q_{ik} \right)^2 q^2_{ik} \left( p_{ij} -r_{il} \right)^2 r^2_{il}}
\\
I_3
&=&
\sum_{ijkl} \int_{pqr}
\frac{q_{ik} \cdot r_{il}}{p^2_{ij} \left( p_{ij} -q_{ik} \right)^2 q^2_{ik} \left( p_{ij} -r_{il} \right)^2 r^2_{il}}.
\end{eqnarray}
Our task is now to compute these master integrals, as well as $I^{\left(3g \right)}$. 

The sum-integral $I_2$ without the holonomy is called $I_{\rm ball}$ in the literature.  Unlike the case of trivial holonomy, the Feynman diagrams (3b)-(3e) cannot be reduced  to a single sum-integral $I_2$. This is because, in a theory with a nontrivial holonomy, the color flow is not fully correlated with the momentum flow. 
If the holonomy is trivial $\theta=0$,  $I_3$ can be reduced to $I_2$, and the results above agree with those obtained  in \cite{Arnold:1994ps}. 

\section{Conclusions and Outlook}
\label{sec:Outlook}
We partially computed the effective potential for the Polyakov loop $V_{\rm free}$ at order $g^3$ and $g^4$ in the Feynman gauge. 
To proceed further, the results of this paper have to be complemented by  $V_{\rm insert}$ (\ref{V_total}), which originates from the delta function constraints of the Polyakov loop.  Then the gauge invariance has to be checked explicitly as a test of the consistency of the calculation. We were able to compute most sum-integrals for $V_{\rm insert}$ at three-loop order using the Poisson summation formula.

Computing the spatial momentum integral of the master integrals at $g^4$ is most efficient in configuration space, see  \cite{Arnold:1994ps}.  We were able to evaluate $I_2$ using the configuration-space technique for spatial momentum and the Poisson resummation formula for the Matsubara sum.  To compute $I_3$ and $I^{(3g)}$ a more elaborated  approach might be required.  This is work in progress.

\end{document}